\documentstyle[aps,preprint]{revtex}
\input{epsf}
\newcommand{\journal}[4]{{{\sl #1}} {\bf #2}, {#3} (#4)}
\newcommand{\mprb}[3]{\journal{Phys.~Rev.~B}{#1}{#2}{#3}}

\newcommand{\mpre}[3]{\journal{Phys.~Rev.~E}{#1}{#2}{#3}}

\def\b{\beta}

\begin{document}
\draft

\title{Effective Field Theory of the Zero-Temperature Triangular-Lattice
Antiferromagnet: A Monte Carlo Study}
\author{Hui Yin and Bulbul Chakraborty}
\address{
The Martin Fisher School of Physics\\
Brandeis University\\
Waltham, MA 02254, USA}
\author{Nicholas Gross}
\address{
College of General Studies\\
Boston University\\
Boston, MA 02215,USA}

\date{\today}
\maketitle

\begin{abstract}
Using a Monte Carlo coarse-graining technique introduced by
Binder~\textit{et~al.}, we have explicitly constructed the continuum field
theory for the zero-temperature triangular Ising antiferromagnet. We verify the
conjecture that this is a gaussian theory of the height variable in the
interface representation of the spin model. We also measure the height-height
correlation function and deduce the stiffness constant. In addition, we
investigate the nature of defect-defect interactions at finite temperatures, and
find that the two-dimensional Coulomb gas scenario applies at low temperatures.
\end{abstract}
\vspace{0.2in}
\pacs{PACS numbers: 05.70.Jk, 05.50.+q, 05.10.Ln, 02.70.Lq}

\section{Introduction}

In recent years, there has been considerable interest in the study of classical
spin systems with critical ground states.  One of the best studied of these is
the triangular-lattice Ising antiferromagnet (TIAFM)\cite{Wannier}.  The critical behavior of these models
can be understood based on an interface representation and an ``effective''
field theory which is gaussian in the height variable\cite{Nienhuis,blote1993}.  In this paper, we
present an explicit construction of the effective field theory based on the
study of Monte Carlo cell distribution functions\cite{Binder1981} of the TIAFM. To accomplish this, a height 
mapping\cite{blote1982} is applied to the system and this variable is then coarse grained to obtain a continuum field theory. Our interest in this model
was stimulated by the observation of anomalously slow dynamics in the compressible
 TIAFM\cite{Leigu1}.  In that model, the supercooled state exhibits an ergodicity-breaking transition which is reminiscent of the structural-glass transition.  This glassy behavior is believed 
to arise from the interaction between strings and vortices\cite{trieste}, the
topological defects present in these models at finite temperature.  In the pure
 TIAFM model, the gaussian theory implies that the vortices interact as charges of a 2-D Coulomb gas\cite{Nienbook}. 
We have analyzed the defect-defect
correlation function at finite temperatures to investigate the nature of the
defect interactions.  The 2-D simulations are consistent with the Coulomb gas
picture at low densities of defects.  

The motivation behind this numerical study was (1) to test the Monte Carlo cell
distribution function technique in a model where the effective field theory is
 well established and (2) to establish a framework for the construction
of effective field theories and effective defect-defect interactions in models
such as the compressible  TIAFM where no adequate field
theory description exists.

The paper is organized as follows.  In section II we describe the cell
distribution function technique.  In section III we present results from the
coarse grained free-energy functional and compare our results to the gaussian
theory.  In section IV we verify the gaussian theory from a 
study of the height-height correlation function and in section V we describe
the study of the defect-defect correlation functions.  Section VI presents our
conclusions and directions for future work.

\section{Cell Distribution Functions and Coarse Grained Free-energy functionals}

Binder and coworkers\cite{Binder1983,Kaski} have studied cell distribution
functions of the three dimensional Ising model on a cubic lattice and constructed coarse-grained Ginzburg-Landau Hamiltonians. In this section, we review this technique and present the results of its application to the triangular Ising ferromagnet as an example. We will describe the application of this technique to the zero-temperature  TIAFM in the next section.

Given a microscopic Hamiltonian such as the Ising model, Monte Carlo(MC) methods can 
be used to sample the distribution functions
$P_L(\lbrace s_i \rbrace)$ of the coarse grained variables 
$$
s_{i}={1\over{L^{2}}}\sum_{l\in{i^{th}}}S_{l}~,
$$
where $S_l$ are the original microscopic spin variables, such as the Ising spins on the 
original lattice, and L is the cell size.
The distributions $P_L(\lbrace s_i \rbrace)$ are assumed to be of the form $e^{-\b \mathcal{F}}$ with $\mathcal{F}$ having the Ginzburg Landau(GL) form in terms of the coarse-grained variables $\{s_{i}\}$. This assumption is expected to be valid when $L>>a$(lattice spacing)
but much smaller than the correlation length, such that the coarse grained variables
do not fluctuate rapidly from cell to cell.  If one could sample the total
distribution function in MC simulations, then this connection could be exploited for the explicit
construction of the GL Hamiltonian by simulating the microscopic model.  Sampling the total distribution function is
essentially an impossible task and therefore, we have followed Binder\cite{Kaski} in studying 
the two simplest reduced distribution functions, the single cell and the joint, nearest-neighbor, two-cell distribution functions which are then parameterized by the GL
form.    To illustrate 
how the cell distribution function method works, we briefly describe its application to the triangular Ising ferromagnet in zero magnetic field, where we choose
the model system size to be 80x80 with periodic boundary conditions and the cell
sizes L to be 4, 8, 10, 16, 20 and 40.

The microscopic Hamiltonian is:
$$
{\mathcal{H}}_{Ising}=-J\sum_{\stackrel{l\not=l'}{<ll'>}}S_{l}S_{l'}
$$
The total distribution function $P_{L}(\{s_{i}\})$ is assumed to be:
\begin{equation}
P_{L}(\{s_{i}\})={1 \over Z} e^{-{\mathcal{F}}_{GL}(\{s_{i}\})}
\end{equation}
where $Z$ is the partition function and the GL Hamiltonian has the form 
\begin{equation}
{\mathcal{F}}_{GL}(\{s_{i}\})=\sum_{i}(\bar{r}_{L}s_{i}^{2}+\bar{u}_{L}s_{i}^4)+\sum_{<ij>}\bar{c}_{L}(s_{i}-s_{j})^2~.
\label{GL}
\end{equation}
The two reduced distribution functions, which are amenable to numerical
calculations, are the two-cell joint distribution function $P_{L}(s_{i},s_{j})$
and the single-cell distribution function $P_{L}(s_{i})$;
\begin{equation}
P_{L}(s_{i},s_{j})=\int \prod_{\stackrel{l \not =i}{\stackrel{l \not =j}{<ij>}}} ds_{l}P_{L}(\{s_{l}\})
\end{equation}
\begin{equation}
P_{L}(s_{i})=\int P_{L}(s_{i},s_{j}) ds_{j}
\end{equation}

The parameterization of the effective Hamiltonian is accomplished by sampling $P_{L}(s_{i},s_{j})$ and $P_{L}(s_{i})$
\begin{equation}
P_{L}(s_{i},s_{j})={1 \over Z'} \exp \{-(c_{L}(s_{i}-s_{j})^2+V_{L}(s_{i})+V_{L}(s_{j}))\}
\label{joint1}
\end{equation}

\begin{equation}
P_{L}(s_{i})={1 \over Z''} e^{-V_{L}(s_{i})}
\label{single}
\end{equation}

where  $Z'$ and $Z''$ are normalization factors, and $$V_{L}(s_{i})=-r_{L}s_{i}^{2}+u_{L}s_{i}^{4}$$

\begin{figure}[h]
\epsfxsize=4.5in \epsfysize=4.5in
\epsfbox{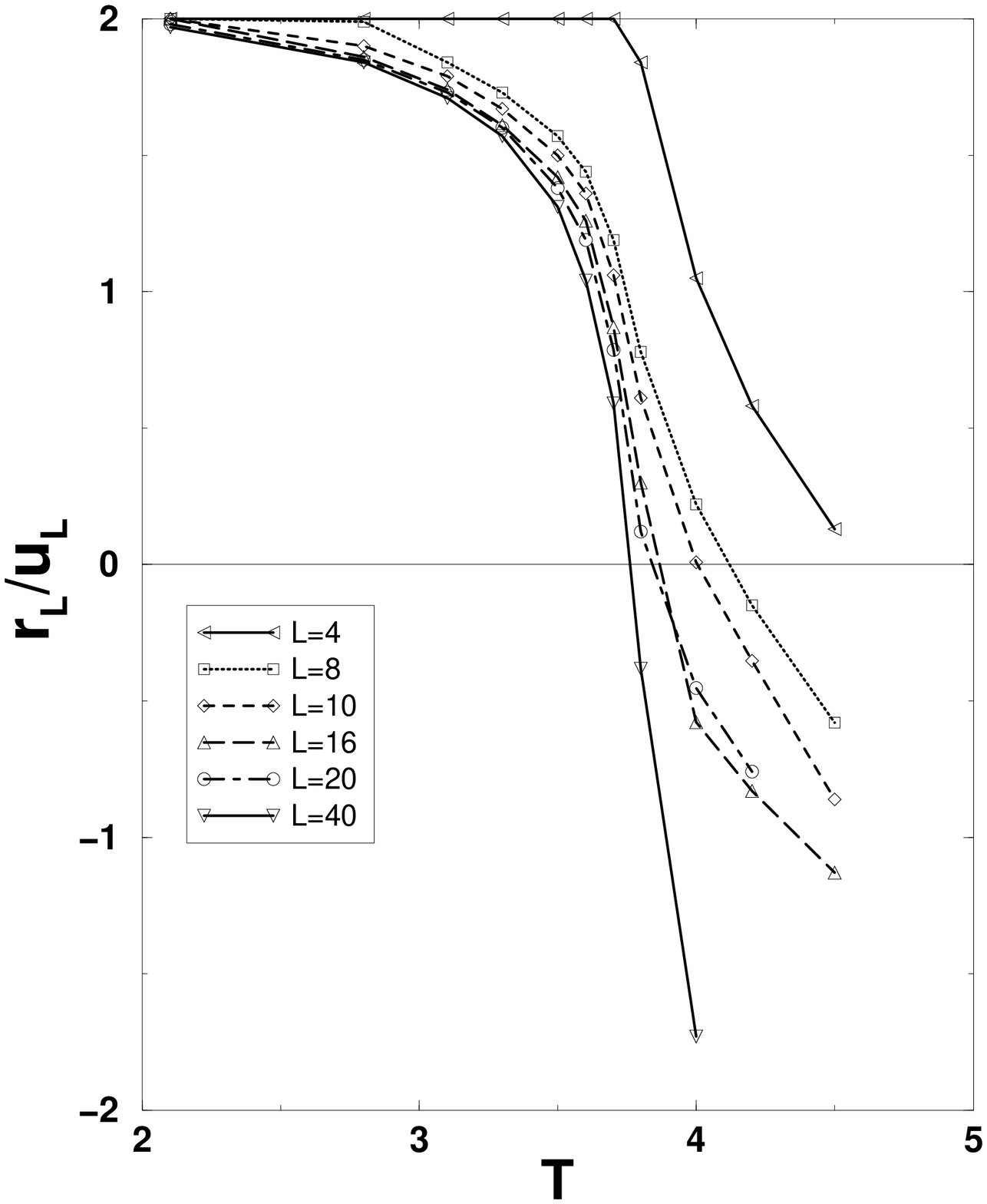}
\vspace{0.3in}
\caption{The ratio${r_{L}(T)\over{u_{L}(T)}}$, calculated from the joint distribution function for different cell sizes. $T_{c}(L)$ is identified as the point where ${r_{L}\over{u_{L}}}=0$. The exact result for $T_{c}(\infty)$ is 3.6 [11]. }
\label{fig1}	
\end{figure}

\begin{figure}[h]
\epsfxsize=4.5in \epsfysize=4.5in
\epsfbox{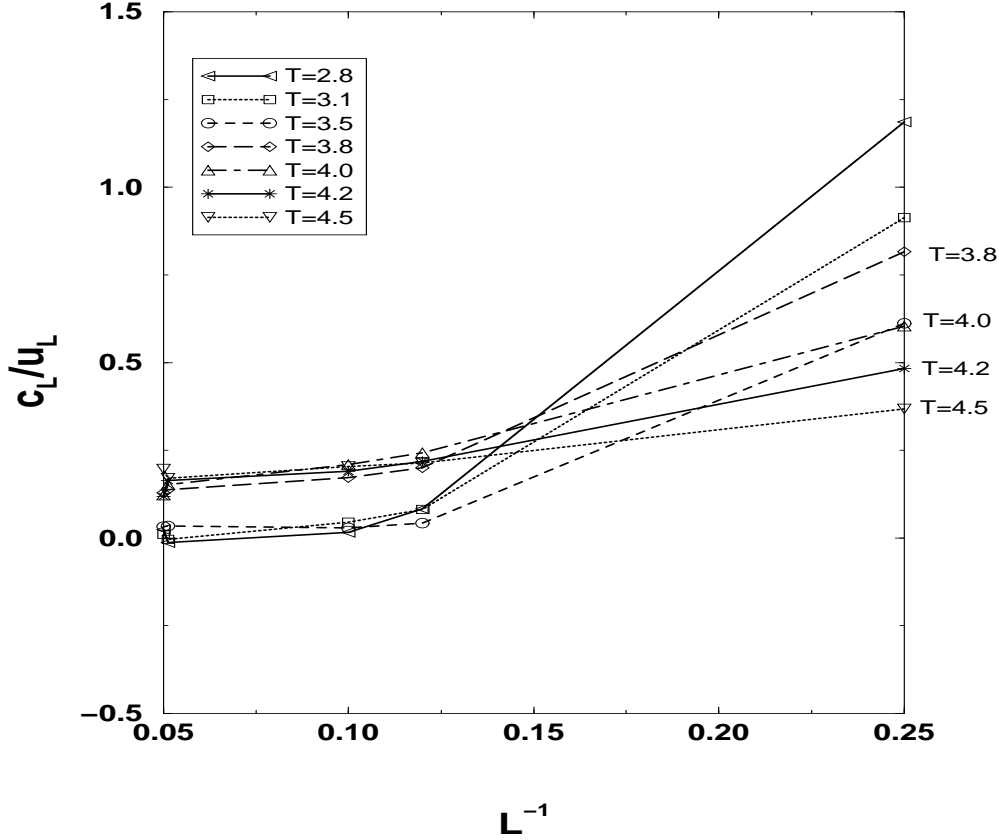}	
\vspace{0.4in}
\caption{The ratio${c_{L}(T)\over{u_{L}(T)}}$, calculated from the joint
distribution function for different temperatures, plotted as a function of
$L^{-1}$. The curves can be divided into two groups. For those with temperatures
higher than $T_{c}(\infty)$=3.6 (identified in the figure), the curves converge to a finite value. For those with temperatures lower than $T_{c}(\infty)$, the curves converge to zero. }
\label{fig2}	
\end{figure}

The $r_{L}$ and $u_{L}$ obtained from $P_{L}(s_{i},s_{j})$ and $P_{L}(s_{i})$ respectively,
can be different since the effect of the gradient term has been integrated over in the single-cell distribution function.  Similar considerations imply that these coefficients  can be different from $\bar{r}_{L}$,
$\bar{u}_{L}$ and $\bar{c}_{L}$ in  Eq.~\ref{GL}. For small values of  $\bar {c_{L}}$, the coupling between different sites is small and all of these coefficients are expected to be approximately equal\cite{Kaski}.
The coefficients $r_{L}$, $u_{L}$ and $c_{L}$, as functions of temperature (T)
and cell size (L),  can be estimated by fitting the measured distribution
function  $P_{L}(s_{i},s_{j})$ to the model
form (Eq.~\ref{joint1}). It is well known that a temperature-driven second order transition exists for this Ising model and the exact result for $T_c(\infty)=3.6$\cite{Baxter}. Fig.~\ref{fig1} shows  ${r_{L}(T)\over{u_{L}(T)}}$ as a
function of  temperature for different cell sizes. There is clear evidence for a 
temperature-driven second order transition at $T_{c}(L)$ where the sign of the
ratio changes.  The L dependence of $T_{c}(L)$ agrees with finite size scaling  
predictions\cite{Binder1981}. The variation of ${c_{L}(T)\over{u_{L}(T)}}$  with
cell size L  (Fig.~\ref{fig2}) shows temperature-dependent behavior as the size
of the cells  approach infinity. For temperatures above $T_{c}({\infty})$, the ratio
approaches a  non-zero fixed point as L approaches infinity. For all other
temperatures, this  ratio approaches zero. The sampling of the single-cell
distribution function  $P_{L}(s_{i})$ (Eq.~\ref{single}) leads to similar
results for $r_{L}$ and  $u_{L}$, which is consistent with the small value
obtained for
$c_{L}$. These results support the assumptions that went into the construction of 
${\mathcal{F}}_{GL}(\{s_{i}\})$.

The results of this section  will be contrasted with
the frustrated antiferromagnetic case in the
next section.  The application of the distribution-function technique to the TIAFM is based on the mapping to an interface model and height
variables.  The coarse graining of these height variables and the GL
parameterization are  described in the next section.
\section{Results of coarse graining and comparison with the gaussian theory}
\subsection{Height Variables and coarse graining at T=0}

One interesting property of the  TIAFM is that the
ground state ensemble  has a one-to-one mapping onto an interface model 
\cite{Nienhuis,blote1993,blote1982}. This provides a simple way of studying the
properties of zero-temperature TIAFM.

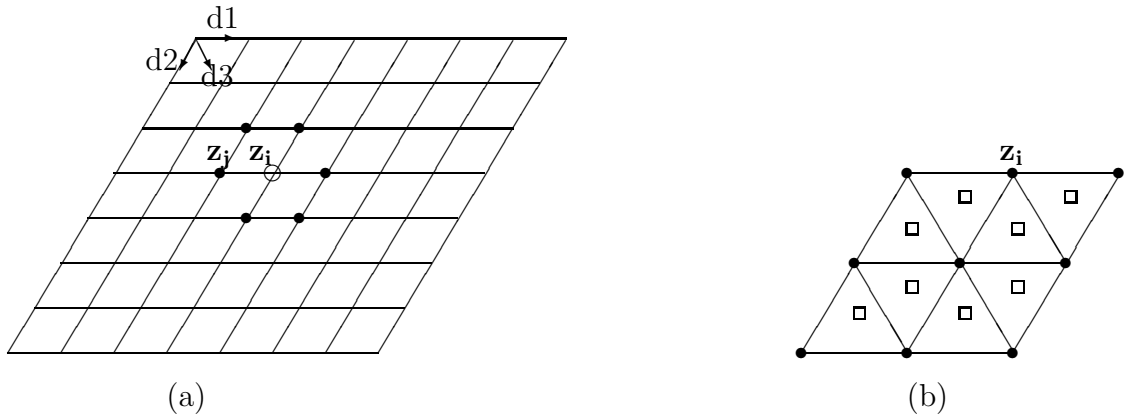
\begin{figure}[h]
\begin{picture}(300,300)(10,10)
\put (20,20) {\line(20,0){140}}
\put (30,37) {\line(20,0){140}}
\put (40,54) {\line(20,0){140}}
\put (50,71) {\line(20,0){140}}
\put (60,88) {\line(20,0){140}}
\put (71,105) {\line(20,0){140}}
\put (81,122) {\line(20,0){140}}
\put (91,139) {\line(20,0){140}}
\put (91,139) {\vector(1,0){15}}
\put (95,143) {d1}
\put (91,139) {\vector(-1,-2){6}}
\put (72,127) {d2}
\put (91,139) {\vector(1,-2){6}}
\put (93,121) {d3}
\put (20,20) {\line(3,5){71}}
\put (40,20) {\line(3,5){71}}
\put (60,20) {\line(3,5){71}}
\put (80,20) {\line(3,5){71}}
\put (100,20) {\line(3,5){71}}
\put (120,20) {\line(3,5){71}}
\put (140,20) {\line(3,5){71}}
\put (160,20) {\line(3,5){71}}
\put (100,88) {\circle*{4}}
\put (95,93) {$\mathbf{z_j}$}
\put (120,88){\circle{6}}
\put (111,93) {$\mathbf{z_i}$}
\put (140,88){\circle*{4}}
\put (110,71){\circle*{4}}
\put (130,71){\circle*{4}}
\put (110,105){\circle*{4}}
\put (130,105){\circle*{4}}
\put (320,20) {\line(20,0){80}}
\put (340,54) {\line(20,0){80}}
\put (360,88) {\line(20,0){80}}
\put (320,20) {\line(3,5){41}}
\put (360,20) {\line(3,5) {41}}
\put (400,20){\line(3,5) {41}}
\put (360,88) {\line(3,-5){40}}
\put (340,54) {\line(3,-5){20}}
\put (400,88){\line(3,-5) {20}}
\put (400,88) {\circle*{4}}
\put (440,88){\circle*{4}}
\put (360,88){\circle*{4}}
\put (340,54){\circle*{4}}
\put (380,54){\circle*{4}}
\put (420,54){\circle*{4}}
\put (320,20){\circle*{4}}
\put (360,20){\circle*{4}}
\put (400,20){\circle*{4}}
\put (420,77){\framebox(4,4){}}
\put (380,77){\framebox(4,4){}}
\put (400,65){\framebox(4,4){}}
\put (360,65){\framebox(4,4){}}
\put (340,33){\framebox(4,4){}}
\put (380,33){\framebox(4,4){}}
\put (360,43){\framebox(4,4){}}
\put (400,43){\framebox(4,4){}}
\put (395,93) {$\mathbf{z_i}$}
\put (80,0) {(a)}
\put (360,0) {(b)}
\end{picture}
\vspace{0.5in}
\caption{(a) The three nearest-neighbor directions d1, d2, d3, and the 6 nearest
neighbors(black dots) of  discrete height variables $z_{i}$(open circle) mapped
onto the rhombus-shaped lattice  used in the simulation. (b) The coarse graining 
cell of size $L=2$. The vertices(black  dots) of lattice are $\{z_{i}\}$ and the
centers of the triangular plaquettes (open square) are $\{H_{\mu}\}$, each of
which is defined as the average of  the three surrounding $z_i$ variables. The
coarse-grained height variable  $h_l$ is defined as the average of all the $H_{\mu}$ variables within this cell.}
\label{lattice}
\end{figure}

The discrete variables in the interface model are height variables \{$z_{l}$\},
which are defined on each site of the triangular lattice. The mapping from the TIAFM to an interface model is realized
by the mapping from spin variables \{$S_{l}$\} to height variables \{$z_{l}$\}. Specifying the height variable
to be zero at a chosen site, the
 following rules (Fig.~\ref{lattice}), provide a unique mapping from \{$S_{l}$\} to \{$z_{l}$\}:
\begin{description}
\item[(1)] Along d1 direction, $\Delta{z}=-1$ for opposite spins and $\Delta{z}=+2$ for same spins.
\item[(2)] Along d2 direction, $\Delta{z}=-1$ for opposite spins and $\Delta{z}=+2$ for same spins.
\item[(3)]Along d3 direction, $\Delta{z}=+1$ for opposite spins and $\Delta{z}=-2$ for same spins.
\end{description}
These height assignments are unique up to the choice of origin as long as we
restrict ourselves to Ising configurations in the ground state ensemble, which
is the set of states that do not have any completely frustrated plaquettes (all
spins same). The microscopic Hamiltonian in term of these height variabls \{$z_{l}$\} can be written directly from the mapping as\cite{Nienhuis} 
$$
{\mathcal{H}}\{z_{l}\}=-J\sum_{<ij>}(3-2|z_i-z_j|)
$$ 
The symmetries of the original Hamiltonian are reflected in
${\mathcal{H}}\{z_{l}\}$. These include Ising up-down symmetry, and sub-lattice symmetry. In
addition, ${\mathcal{H}}\{z_{l}\}$ has a symmetry with respect to global, discrete shifts of
the height  variable. A detailed discussion of these symmetries can be found in Bl\"{o}te
\textit{et al.}\cite{blote1993}. An interesting property of this mapping is that
for any ground state the height variable  modulo 3 is the same for all spins on
the same  sub-lattice.

 Starting from \{$z_{l}$\}, we can define a new set of height variables
\{$H_{\mu}$\} which are situated on the dual lattice, by averaging  the three neighboring $z_{l}$ on each triangular
plaquette: $H_{\mu}=\frac{1}{3} (z_{1}+z_{2}+z_{3})$ where $z_{1},z_{2}$,
and $z_{3}$ are  on the three vertices of the ${\mu}^{th}$ plaquette\cite{Zeng}
as shown in Fig.~\ref{lattice}. Since the sum
of $z_{1},z_{2}$,  and $z_{3}$ are multiples of 3, the \{$H_{\mu}$\}
is a set of  integers.

We apply the coarse graining technique to the height variables \{$H_{\mu}$\}.  The
coarse-graining  cell is chosen to be a rhombus with the linear size L and the
coarse-grained  height variables \{$h_{i}$\} are defined as the average of all
the  $H_{\mu}$  within the cell.
$$
h_{i}={1\over N}\sum_{{\mu}\in{i^{th}}}H_{\mu} 
$$ 
where N is the total
number of $H_{\mu}$  variables within the $i^{th}$  cell. The coarse-graining cell is depicted schematically in Fig.~\ref{lattice}. 
As L approaches infinity, the $\{h_{i}\}$ become continuous variables. The
effective field theory is  based on these continuous variables $\{h_{i}\}$,
which describe the  roughness of the interface. In the rough phase of the
interface model, the  average tilt of the interface, which is defined as the
difference between  $<h_{m}>$ and $<h_{n}>$ with $m^{th}$ cell and $n^{th}$ cell
separated by the  system size, is zero. For an arbitrary spin configuration, a tilt can be frozen in\cite{blote1982}. We will avoid such
configurations in our  simulations. Also \{$<h_{i}>$\} may be non-zero for a
general choice of  height origin and the Hamiltonian should only depend on the
fluctuation  $h_{i}-<h_{i}>$\cite{blote1982}.

We performed Monte Carlo simulations on a system of size 600x600 shaped as a
rhombus and looked at cell sizes 10, 15, 20, 30, 40, 50, 60, 75, 100. We set
the initial  configuration to be one of the flat states\cite{Zeng}, where
$\{H_{\mu}\}$ is uniform in space. The dynamics we use in the simulation is single spin-flip Metropolis algorithm at zero temperature, where the energy is not allowed to increase and thus not all spins are flippable. The first 20,000 configurations are discarded to ensure that
measurements are taken in equilibrium. The sampling is done every ten MC
steps and for a  total of 6,000 configurations. Since we use single
spin-flip MC dynamics,  the average tilt is unchanged from its initial value
of zero during  the MC runs\cite{Hoogland}. In the following sections, we will
use $h_{i}$ to  represent $h_{i}-<h_{i}>$. 

\subsection{Results and comparison with the gaussian theory}
It has been conjectured\cite{Nienhuis,blote1993} that in terms of the height variables,
the continuum  theory for the zero temperature TIAFM is gaussian:
\begin{equation}
{\mathcal{F}}(h(\vec{r}))=\int\ d\vec{r}\ \frac{\bar{c}}{2} \ (\nabla h(\vec{r}))^2
\label{continuum}
\end{equation}
with a value of $\bar{c}=\pi/9$. This stiffness constant was identified by
Bl\"{o}te \textit{et al} through the  correspondence between the exact
calculation of the spin-spin correlation  function in the spin
model\cite{Stephenson} and the discrete  height-height correlation function in
real space in the interface model\cite{blote1982}. The stiffness constant was
also obtained numerically by Zeng and Henley from measurements of the discrete
height correlation function in Fourier space\cite{Zeng}. In this section, we
deduce the stiffness constant by an explicit construction of the coarse-grained
free energy.

 Assuming a Ginzburg-Landau form for the Hamiltonian in terms of the continuous height variables, 
\begin{equation}
{\mathcal{F}}(h(\vec{r}))=\int\ d\vec{r}\ \{\frac{\bar{a}}{2}\ h(\vec{r})^2+\frac{\bar{c}}{2} \ (\nabla h(\vec{r}))^2\}
\label{assumption}
\end{equation}
we proceed to obtain $\bar{a}$ and $\bar{c}$ from our simulations by parameterizing the joint distribution function of the coarse-grained height variables $P_{L}(h_{i},h_{j})$. 

 
As in the triangular ferromagnet, we study the nearest-neighbor joint
distribution function of heights,
\begin{equation}
P_{L}(h_{i},h_{j})=\int \prod_{\stackrel{l \not =i}{\stackrel{l \not =j}{<ij>}}} 
dh_{l}P_{L}(\{h_{l}\})
\end{equation}
where $P_{L}(\{h_{l}\})$ is the total cell distribution function. We
parameterize the joint  distribution function as 

\begin{equation}
\label{eq:joint}
P_{L}(h_{i},h_{j})={1 \over Z'}\exp \{-[c_{L}(h_{i}-h_{j})^2+a_{L}(h_{i}^{2}+h_{j}^{2})]\}
\end{equation}
where $Z'$ is a normalization factor.

Anticipating a critical point at T=0, where $a_{L}$ $\rightarrow$ 0, we have
neglected the  quartic term in Eq.~\ref{eq:joint}. This will be justified a
posteriori from our  numerical study. From the joint distribution function, we
can extract  $a_{L}$ and $c_{L}$ as in the ferromagnetic Ising model.

\begin{figure}[h]
\epsfxsize=4.5in \epsfysize=4.5in
\epsfbox{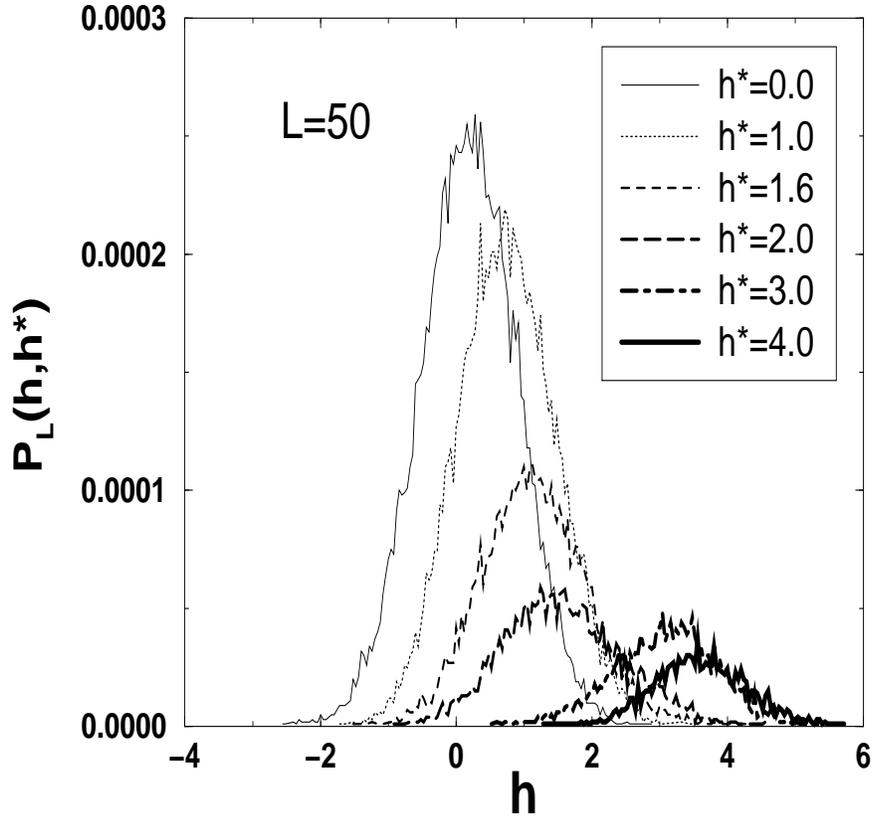}
\vspace{0.5in}
\caption{The cuts of $P_L(h,h*)$ for cell size L=50 and different values of h*,
demonstrating the dominance of  the gradient term(\textit{cf.} text) .}
\label{figcuts}
\end{figure}

We examine a typical $P_{L}(h,h^{*})$ for cell size L=50. One-dimensional
cuts of  $P_{L}(h_{i},h_{j})$ at different fixed values $h_{j}=h^{*}$ are shown
in  Fig.~\ref{figcuts}. The peak and shape of the cuts can be deduced from the
distribution  function (Eq.~\ref{eq:joint}) which is a product of two Gaussians, $\exp (-a_{L}h^2)$  and $\exp (-c_{L}(h-h^{*})^2)$.  When $h^{*}$ differs from 0 by a large amount, the plots in
Fig.~\ref{figcuts}  show that the actual peak of the cut is much closer to
$h^{*}$ than to 0,  which implies that the $\exp (-c_{L}(h-h^{*})^2)$ is the
dominant term  in  the product of Gaussians and the $\exp (-a_{L}h^2)$ term acts
more as a  pre-factor modulating the amplitude of the peak. The width of the cut
is also seen to  be determined mainly by $c_{L}$. Fig.~\ref{figcuts} indicates
that $a_{L}$ is  small compared with $c_{L}$. We also analyze the importance of
higher order  gradient terms by measuring the non-gaussian parameter ($g_{L}$)
in the  distribution of $v={1 \over 2} (h_{i}-h_{j})$:
\begin{equation}
g_{L}\equiv1-{<v^4>_{L} \over 3<v^2>^2_{L}}
\end{equation}

We find $g_{L}$ to be much smaller than 1, which is consistent with the assumed
form of the gradient term in the joint  distribution function. Extracting quantitative
information about the  coefficients $a_{L}$ and $c_{L}$ from fits to the
distribution functions,  turns out to be difficult because of the essential
two-dimensional  nature of the distribution function
$P_{L}(h_{i},h_{j})$($c_{L}$ is large).  Instead, we resort to measurement of moments for  extracting quantitative information.
 
The parameters $a_{L}$ and $c_{L}$ can be related to the moments of
$\{h_{i}\}$. We have  calculated various moments of these coarse-grained
variables directly from  the MC simulations. In general, the parameterization of
the joint  distribution function (such as Eq.~\ref{eq:joint}) provides a
connection between  these moments and the coefficients of the GL
Hamiltonian. For the model  of Eq.~\ref{eq:joint}, the relation between $a_{L}$
and $c_{L}$ and  the moments can be shown to be:

\begin{equation}	
a_{L}={1 \over 2(<h^2>+<h_{1}h_{2}>)}
\label{al}
\end{equation}

\begin{equation}
c_{L}={<h_{1}h_{2}>\over 2(<h^{2}>+<h_{1}h_{2}>)(<h^{2}>-<h_{1}h_{2}>)} 
\label{cl}
\end{equation}

implying that
\begin{equation}
\label{ratio}
\frac{c_{L}}{a_{L}}={<h_{1}h_{2}>\over(<h^{2}>-<h_{1}h_{2}>)}
\end{equation}

\begin{figure}[h]
\epsfxsize=4.5in \epsfysize=4.5in
\epsfbox{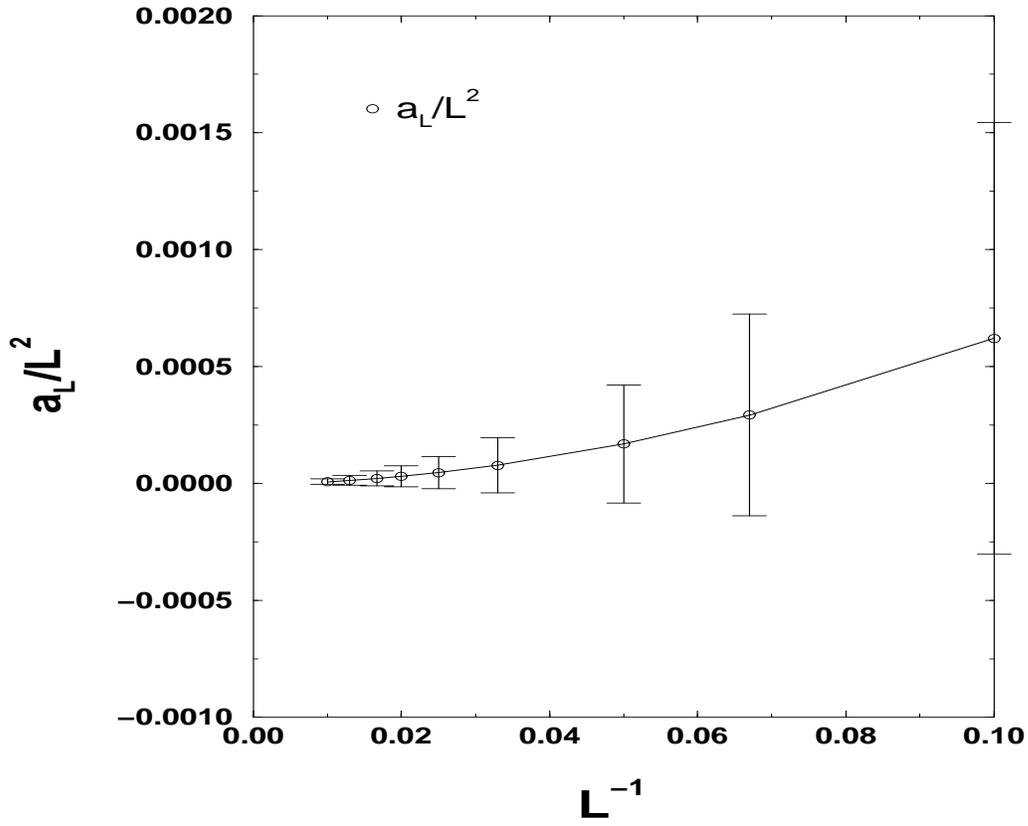}
\vspace{0.3in}
\caption{Plots of $\bar{a}_{L}=a_{L}/L^{2}$ for different cell sizes. The open
circles are the data points and  the solid line is the second degree
polynomial fit. The error  bars are obtained from the standard deviations
calculated from appropriate higher  moments of $h_{i}$ and $h_{j}$. The fitting
of $a_{L}/L^{2}$ leads to a  constant term with a value of 6.183E-6, which
implies that the quadratic term in the  field theory $\bar{a}$
(cf.~Eq.~\ref{a_bar}) becomes 
vanishingly small as  $L\rightarrow$$\infty$.}
\label{fig:al_L2}
\end{figure}

\begin{figure}[h]
\epsfxsize=4.5in \epsfysize=4.5in
\epsfbox{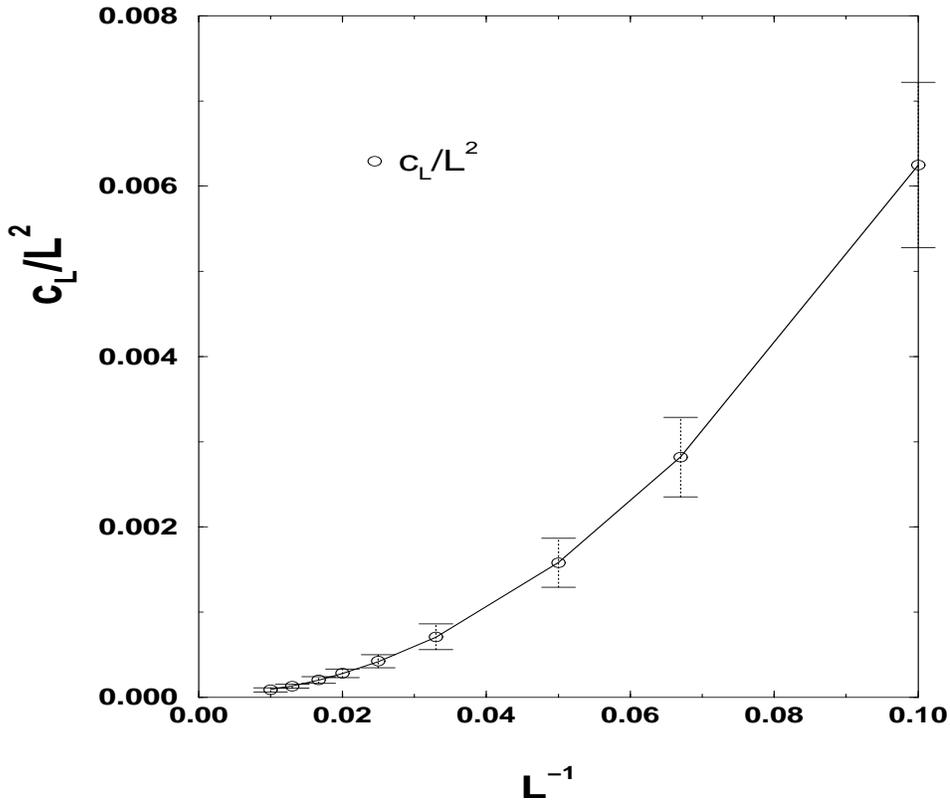}
\vspace{0.3in}
\caption{Plots of $c_{L}/L^{2}$ for different cell sizes. The open circles are
the data points and the solid  line is the second degree polynomial
fit. The error bars are the  standard deviations calculated from appropriate
higher moments of $h_{i}$ and  $h_{j}$. The fitting of $c_{L}/L^{2}$ leads to a
quadratic term with a  coefficient 0.622, and vanishingly small constant and
linear terms, which implies that the gradient term
of the continuous  theory(cf.~Eq.~\ref{c_bar}), $\bar{c}$ approaches 0.392 as $L$$\rightarrow$$\infty$.}
\label{fig:cl_L2}
\end{figure}

Here $h_1$ and $h_2$ refer to coarse-grained height variables on
nearest-neighbor cells separated by the coarse-graining cell size $L$.  
These relations, between the parameters $a_L$ and $c_L$ and the moments, also
provide the connection between $a_L$ and $c_L$ and the corresponding parameters $\bar{a}$ and $\bar{c}$ in
continuum field theories such as in Eq.~\ref{assumption}.  Using the known
expressions for the correlations functions in the continuum
model in two dimensions\cite{Goldenfeld}, these relations are:
\begin{equation}
\bar a = 4\,a_{L}\,L^{-2}
\label{a_bar}
\end{equation}
and
\begin{equation}
\bar c = 2\,w_0\,\frac{\ln {(\pi)}}{\pi}\,c_{L}
\label{c_bar}
\end{equation}
The factor $w_0=\sqrt{3}/2$ arises from the volume per unit cell in the triangular
lattice.  The prediction from the gaussian theory is $\bar{a}=0$ and $\bar{c}=
\pi/9$.  
We
expect that, if the  gaussian theory is correct, then $\bar{a}$ should approach zero
and $\bar{c}$  should remain constant as $L$$\rightarrow$$\infty$. Because of the
connection to the  continuum theory(Eq.~\ref{a_bar}), it is natural to fit $a_{L}/L^{2}$, to a second-order polynomial in $1/L$.  The results of the fitting are 
shown in  Fig.~\ref{fig:al_L2}. The value of $\bar{a}$ can be deduced from the
constant term extracted from the fitting and is found to be $6.183*10^{-6}$ implying that the quadratic term in the Ginzburg-Landau Hamiltonian becomes negligible as $L$$\rightarrow$$\infty$. 
It is difficult to extract the value of
$\bar{c}$ from the measured values of $c_{L}$ because of the fact that the errors increase
with L\cite{Error}.  If, instead, we analyze $c_{L}/L^{2}$, the error bars
actually decrease with $1/L$ as seen from Fig~\ref{fig:cl_L2}, and  this provides a
more precise way of determining the value of $\bar{c}$. As shown in
Fig.~\ref{fig:cl_L2}, the data for $c_{L}/L^{2}$ can be fit very well to a
second-order polynomial in $1/L$ with vanishingly small  coefficients of the
constant and linear term in $1/L$.
The value of $c_{L}$ obtained from this fit is $0.622$.  The standard
deviations, calculated from the  appropriate higher moments of $h_{i}$ and
$h_{j}$ based on Eq.~\ref{al} and  Eq.~\ref{cl}, are plotted as error bars. 

These   results for the parameters obtained from our analysis of the moments are 
completely consistent with the qualitative behavior deduced from the joint
distribution functions, $P_{L}(h_{i},h_{j})$, shown in Fig.~\ref{figcuts}.

The results shown in Fig.~\ref{fig:al_L2} and Fig.~\ref{fig:cl_L2}
verify the  gaussian nature of the continuum theory. The numerical value of
$\bar{c}$ extracted by applying Eq.~\ref{c_bar} to the fitted value of $c_{L}$ is 0.392, which is in very good
agreement with 
the  conjectured stiffness constant of $\pi/9=0.349$~\cite{Nienhuis,blote1993}.

\section{Height-Height correlation function}

In this section we will briefly present our calculation of the correlation
function of height variables in  Fourier space, {\it i.e.}, the power spectrum
of the height variable, which can be used to deduce the stiffness constant as in the investigation of Zeng and Henley\cite{Zeng}. In the next  section we further extend this technique to study the
defect-defect  interactions at finite temperatures by measuring the defect density correlation function.

As implied by Eq.~\ref{continuum}, the stiffness constant $\bar{c}$ of the
interface model can be  directly related to the correlation function of height
variables  \{$h_{i}$\} in Fourier space. 
\begin{equation}
{\mathcal{F}}\{h(\vec{r})\}=\int d\vec{r}\ \frac{\bar{c}}{2} \ (\nabla
h(\vec{r}))^{2}=\sum_{\vec{q}}  \frac{\bar{c}}{2} \ |\vec{q}|^{2}\ |h(\vec{q})|^2
\end{equation}

Thus, the correlation function $<|h(\vec{q})|^2>$ is:
\begin{equation}
<|h(\vec{q})|^2>={1 \over \bar{c}\,|\vec{q}|^2}
\end{equation}

The stiffness constant $\bar{c}$ can be extracted from a knowledge of
$<|h(\vec{q})|^2>$.
Numerical measurements of $<|h(\vec{q})|^2>$ are, however, arduous, and instead,
we study the  correlation function of the discrete height variables,
$<|H(\vec{q})|^2>$, which  should resemble $<|h(\vec{q})|^2>$ for small q except for a factor of $\sqrt{3}/2$ coming from the volume per lattice site. Therefore, the $<|H(\vec{q})|^2>$ for small q can be written as:
\begin{equation}
<|H(\vec{q})|^2>={1 \over c\,|\vec{q}|^2}
\end{equation}
where 
\begin{equation}
c=\frac{2}{\sqrt{3}}\,\bar{c}
\label{discrete_c}
\end{equation}
We simulate a system of size 240x240. We start with the same flat
states as the ones used in  the coarse-graining simulations. After discarding the
first 20,000  configurations, the simulations are sampled at every 10 MC steps
for a total of 8,000  configurations. For each sampled configuration we
calculate $|H(\vec{q})|$, the  two-dimensional Fourier transform of the discrete
height variables \{$H_{\mu}$\} (cf section IIIA) and average  $|H(\vec{q})|^2$
over the sampled MC configurations.

\begin{figure}[h]
\epsfxsize=4.5in \epsfysize=4.5in
\epsfbox{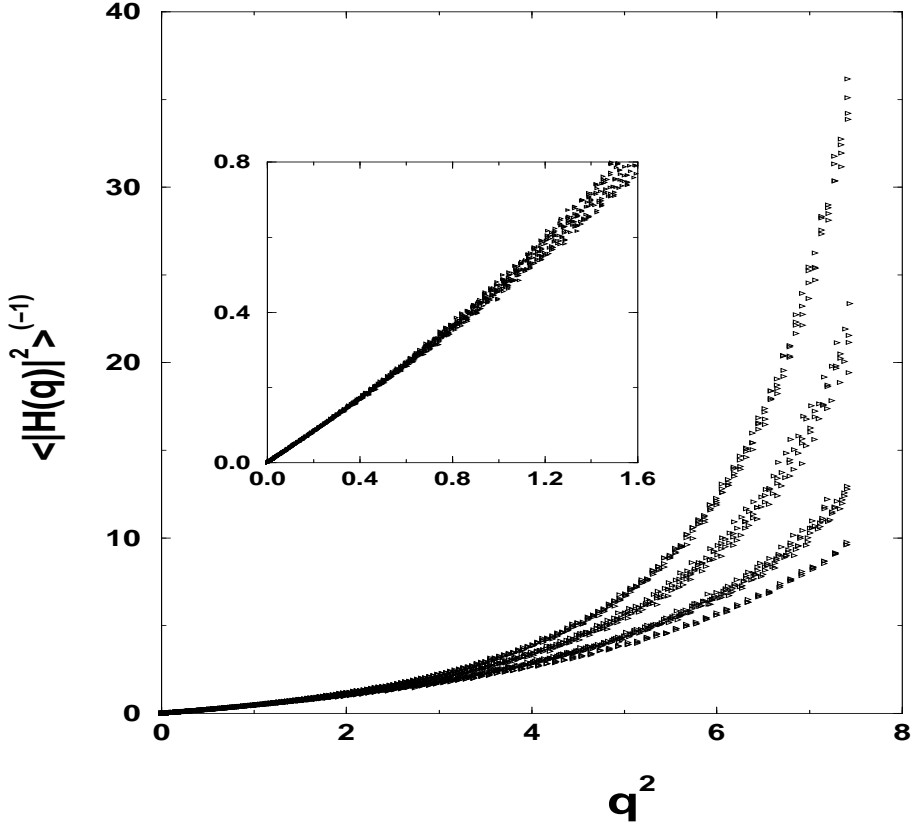}
\vspace{0.3in}
\caption{Plots of $<|H(\vec{q})|^2>^{-1}$ vs. $|\vec{q}|^2$. Different data sets
are from the cuts of  two-dimensonal data along different directions in
q-space. The correlation function shows  significant anisotropy for
$q^2\ge1.6$. The value of $c=0.414\pm0.016$  is obtained from the fitting of
curves in the region $q^2<0.2$ to a second-degree polynomial in $q^2$.  The stiffness constant, $\bar{c}$, deduced from 
this fit is
0.360.}
\label{fig:anisotropic}
\end{figure}

For simplicity, one-dimensonal cuts of $<|H(\vec{q})|^2>^{-1}$ along different
directions in q-space are shown in Fig.~\ref{fig:anisotropic}. As seen from
Fig.~\ref{fig:anisotropic}, different cuts collapse on top of one another only
for small values of $q^2$. This  is due to the fact that the correlation
function is isotropic for small values  of q because of the six-fold rotational
symmetry of the triangular  lattice. As seen from the insert, the anisotropy
starts to be significant when  $q^2>1.6$. We have fitted
$<|H(\vec{q})|^2>^{-1}$ to a second-order polynomial in $q^2$, restricting the
fitting region to  $q^2<0.2$. A value of
$c=0.414\pm0.016$ was extracted from the linear term of the fit.  The non-linear 
terms were small, and an estimate of the error on $c$ was obtained from the
standard deviation of the linear term extracted from fits to different segments
of the curves ($q^2<0.2$). From Eq.~\ref{discrete_c}, the stiffness
constant  can be deduced to be  $\bar{c}=0.360$, compared with 0.392 from
our previous real space  coarse-graining approach and $\frac{\pi}{9}=0.349$ from
Bl\"{o}te \textit{et al}\cite{blote1982}.

The similarity of the results obtained from the real-space analysis and the
power-spectrum analysis is quite remarkable and underlines the strength of these 
numerical techniques.  A comparison of the two different analysis techniques,
however, makes it clear that the power-spectrum analysis is simpler to implement 
and is the more desirable technique.  The question of how well either of these
techniques will fare when applied to a model with general non-linear interactions, where
very little is known analytically, is still open.

\section{Defect Interactions}
At finite temperatures, there are completely frustrated plaquettes which can be
viewed as vortex excitations in the height variable\cite{Nienhuis,blote1993}.  There are
two types of vortices corresponding to the two
different orientations of the 
frustrated triangular plaquettes.  These vortex excitations interact like
charges via a two-dimensional Coulomb potential\cite{blote1993,Nienbook}.  

An alternative to the gaussian description is, therefore, a Coulomb gas
description involving electric and magnetic charges corresponding to spin waves
and vortices, respectively\cite{blote1993,Nienbook}.  This connection suggests that
extracting effective defect-defect interactions could be an alternative to
constructing coarse-grained free energies.  In complicated models and especially 
in analyzing dynamics, this might be the more viable alternative.  We therefore
wanted to numerically extract the effective vortex-vortex interaction in the
TIAFM and verify the Coulomb gas scenario.  

In the regime of low defect-density,
where mean field theory is expected to hold, the effective interactions can be  
related to the density-density correlation
function\cite{Chaikin}.  The appropriate mean-field theory for our system with 
positive and
negative charges is Debye-H\"{u}ckle theory\cite{Chaikin} which predicts that
the charge-density correlation function has the following form:
\begin{equation}
<|\rho(q)|^2>={\kappa^2 \over 4\pi} {q^2 \over q^2+\kappa^2}
\label{defect_corr}
\end{equation}

where 
\begin{equation}
\kappa^2=4\pi(<n_+>+<n_->)
\label{screen}
\end{equation}
is the square of an inverse Debye-H\"{u}ckle screening length. $<n_\pm>$ are the average densities of positive and negative charges and $\rho(q)$=$n_+(q)+n_-(q)$, where $n_\pm(q)$ are the Fourier transform of the real space charge density functions.

If the vortices in our model behave as a Coulomb gas, then we expect to find
this behavior of the charge-density correlation function at low temperatures
where the vortex density is low.  

We studied the  charge density correlation function in a manner similar
to that employed in the previous section for studying the height-height
correlation function. The MC simulations were performed on
a 120x120 system  and after discarding the first 10000 configurations,
samples were taken every 5 configurations. The correlation function,
$<|\rho(q)|^2>$,  was measured at several temperatures. 

\begin{figure}[h]
\epsfxsize=4.5in \epsfysize=4.5in
\epsfbox{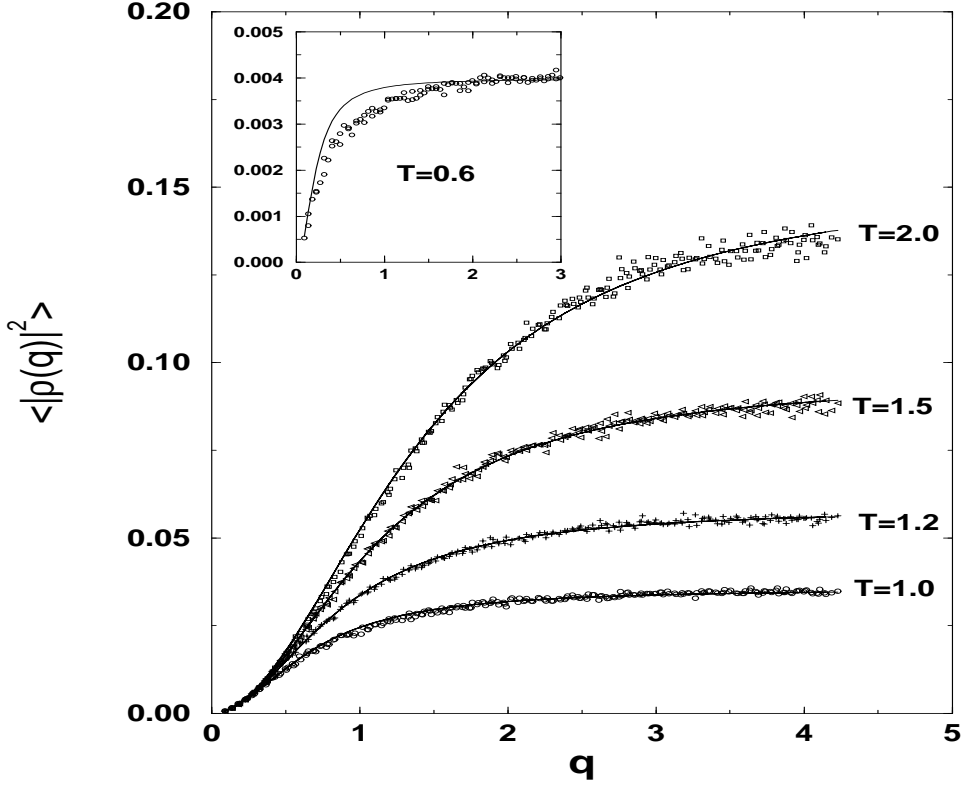}
\vspace{0.3in}
\caption{Plots of the defect density correlation function at T=1.0,
1.2, 1.5, 2.0. The data points are taken from cuts of two-dimensional 
data along one chosen direction since we did not observe any anisotropy ({\it
cf} text), and the
solid lines are from the fittings to a Debye-H\"{u}ckle
form(Eq.~\ref{defect_corr}). The fittings lead to $\kappa^2_{f}$ at
different temperatures. The $\kappa^2_{f}$ and the analytical values
of $\kappa^2_{a}$ extracted from the total defect density for the same
temperature~(cf.~Eq.~\ref{screen}) are shown in
Fig.~\ref{fig:kappa_T}. The total defect density is 0.030, 0.051, 0.085, 0.110,
respectively, for T=1.0, 1.2, 1.5, 2.0. The inset shows the fitting at the
lowest temperature, T=0.6 where the defect density is 0.003 and the results
suffer from a lack of statistics}
\label{fig:defect}
\end{figure}

The simulation results are shown in Fig.~\ref{fig:defect} along with
the fits to the Debye-H\"{u}ckle form(Eq.~\ref{defect_corr}). The figure shows a 
one-dimensional cut of $<|\rho(q)|^2>$ along a particular direction.  In
contrast to the height power spectrum, the defect density power spectrum was found to be 
isotropic within the range of $q$ studied.  
The Debye-H\"{u}ckle theory models the data well for all
temperatures except the lowest temperature, $T=0.6$, which is shown in
the insert. The lowest temperature data is suspect because the defect
density is low and statistics are difficult to obtain.  The isotropy of the
$<|\rho(q)|^2>$ data 
imply that the screening length and, therefore, the interactions between the
defects on the triangular lattice is isotropic down to length scales of the
order of the lattice spacing.

\begin{figure}[h]
\epsfxsize=4.5in \epsfysize=4.5in
\epsfbox{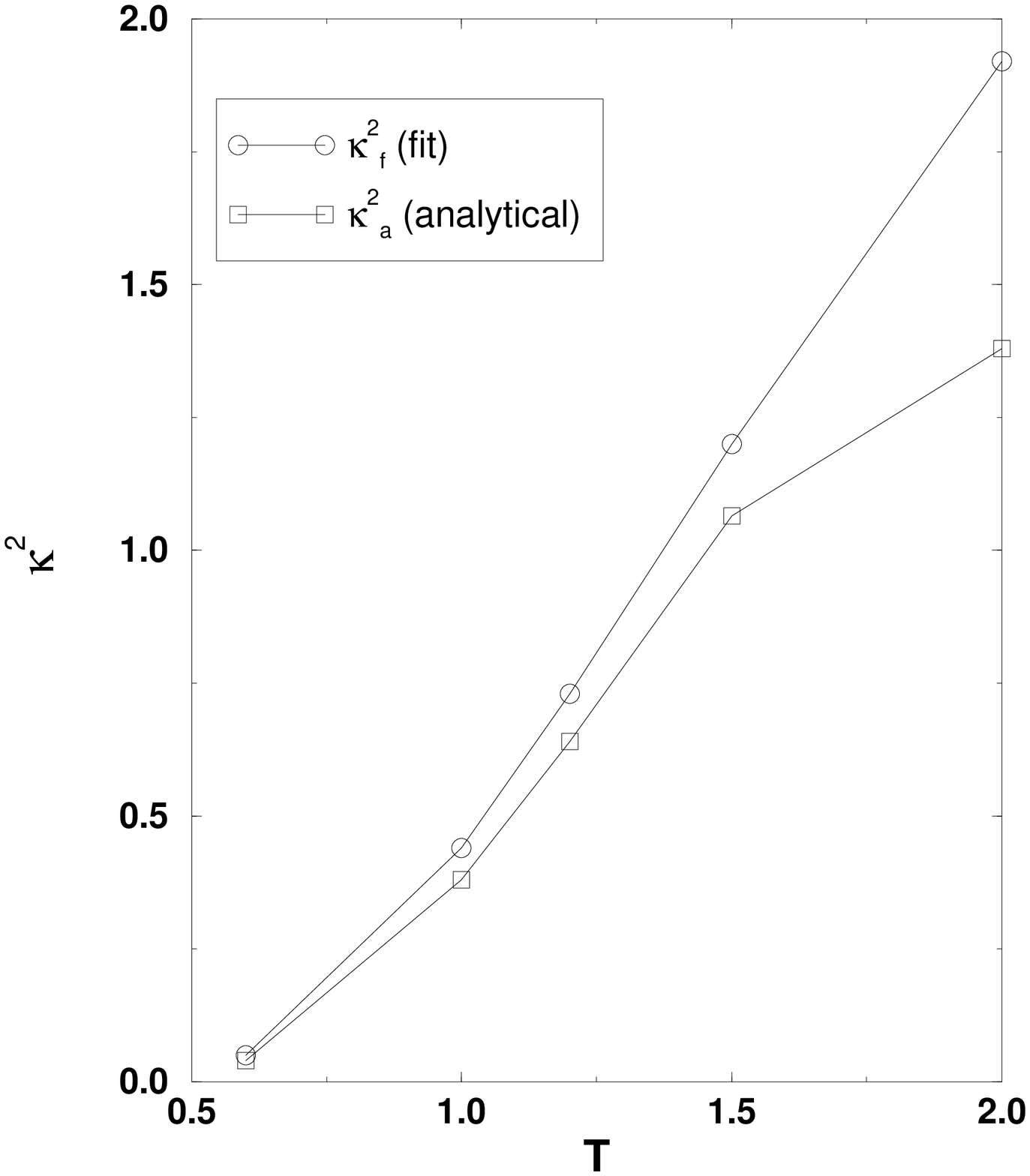}
\vspace{0.3in}
\caption{Plots of $\kappa^2_{f}$, the inverse screening length obtained from fitting compared to $\kappa^2_{a}$, the value obtained from Eq.~\ref{screen}. The  two curves deviate significantly only at the highest
temperature where the defect picture itself starts to break down ({\it cf} text).}
\label{fig:kappa_T}
\end{figure}

From the fits to the Debye-H\"{u}ckle theory the inverse square of the
screening length, $\kappa^2$ can be extracted and compared to the analytic
form given in Eq.~(\ref{screen}) which uses as an input the measured
defect density. In Fig.~\ref{fig:kappa_T}, the fitted values, $\kappa^2_f$ and
the analytic values,  $\kappa^2_a$ are
plotted versus temperature.  Except at the highest temperature, T=2.0, the
agreement between the two different estimates is very good.  Interestingly, the
temperature at which the two  $\kappa$'s differ significantly from each other is 
also where the defect density differs from a simple Arrhenius prediction
corresponding to an activation energy of $4J$, the single-defect creation
energy. The observation of the large defect density at T=2.0,
$<n_+>+<n_->=0.110$, indicates that the defects  are covering the whole lattice
at this temperature and the single-defect picture becomes inappropriate.

These numerical studies show that it is possible to extract effective
defect-defect interactions from measurements of the defect-density power
spectrum and verifies that the interaction, in the TIAFM, is Coulombic.   A
numerical measurement of the force between defects has
been 
made via a study of the dynamics of the defects in TIAFM\cite{defects}.  This
study could not confirm the existence of a Coulomb force from the behavior of
the defect density at long times, however, it was argued that the force
could not be falling of more slowly than $1/r$.

\section{Conclusion}

Based on a mapping to an interface model, we have constructed a coarse-grained
Hamiltonian for the zero-temperature TIAFM  by studying the joint distribution
function of the coarse-grained  height
variables.  Our numerical work confirms that the effective field theory is
gaussian in terms of the  coarse-grained height variables and we obtain a
numerical estimate of the stiffness  constant which is in a good agreement with
the analytical prediction. The  connection between the parameters in joint
distribution function and the moments  of height variables eliminates the
arduous work of directly relating the  parameters of joint distribution function
to those of the total  distribution function.  We also directly measured the
power spectrum of the 
height and obtained a
value of the stiffness  constant,  which is very close to the analytic
prediction and the estimate from the real-space coarse-graining. 
This suggests that measurement of power-spectrum might provide a simple route
towards the construction of effective field theories.  We also demonstrated 
that effective  defect-defect interactions can be extracted from
numerical studies of the defect-density power spectrum..

The compressible TIAFM\cite{kardar,Leigu} is currently
being studied using these techniques.  The effective field theory and
defect-defect  interactions in
this model are of interest for the study of alloys with the elastic 
interactions, and  for the study of glassy dynamics\cite{Leigu1}.

This work was supported in part by the DOE grant DE-FG02-ER45495. We would like
to  thank David Olmsted,  Mike Ignatiev, and Mark Sobkowicz for helpful
discussions, and Jane' Kondev for his insightful comments on the manuscript.

\vspace{0.5in}

\vspace{0.5in}

\vspace{0.5in}

\vspace{0.5in}

\vspace{0.5in}


\begin{thebibliography}{99}
\bibitem{Error} Larger cell size L implies less cells for a given system size, thus less coarse grained height variables $\{h_{i}\}$ to average over for each sampled MC configuration.
\end{thebibliography}
\end{document}